\documentclass[aps,prl,twocolumn,amsmath,amssymb]{revtex4-2}

\usepackage{graphicx}
\usepackage{color}
\usepackage[hidelinks]{hyperref}
\usepackage{orcidlink}

\hypersetup{colorlinks=true, linkcolor=black, citecolor=black, urlcolor=blue}

\begin{document}

\author{Christoph Hotter\,\orcidlink{0009-0003-3854-0264}}
\author{Helmut Ritsch\,\orcidlink{0000-0001-7013-5208}}
\author{Karol Gietka\,\orcidlink{0000-0001-7700-3208}}
\email[]{karol.gietka@uibk.ac.at}

\affiliation{Institut f\"ur Theoretische Physik, Universit\"at Innsbruck, A-6020 Innsbruck, Austria} 

\title{Combining critical and quantum metrology}

%%%%%%%%%%%%%%%%%%%%%%%%%%%%%%%%%%%%%%%%%%%%%%%%%%%%%%%%%%%%%%%%%%%%%%%%%%%%
%%%%%%%%%%%%%%%%%%%%%%%%%%%%%%%  ABSTRACT  %%%%%%%%%%%%%%%%%%%%%%%%%%%%%%%%%
%%%%%%%%%%%%%%%%%%%%%%%%%%%%%%%%%%%%%%%%%%%%%%%%%%%%%%%%%%%%%%%%%%%%%%%%%%%%

\begin{abstract}
Critical metrology relies on the precise preparation of a system in its ground state near a quantum phase transition point where quantum correlations get very strong. Typically this increases the quantum Fisher information with respect to changes in system parameters and thus improves the optimally possible measurement precision limited by the Cramér-Rao bound. Hence critical metrology involves encoding information about the unknown parameter in changes of the system's ground state. Conversely, in conventional metrology methods like Ramsey interferometry, the eigenstates of the system remain unchanged, and information about the unknown parameter is encoded in the relative phases that excited system states accumulate during their time evolution. Here we introduce an approach combining these two methodologies into a unified protocol applicable to closed and driven-dissipative systems. We show that the quantum Fisher information in this case exhibits an additional interference term originating from the interplay between eigenstate and relative phase changes.  We provide analytical expressions for the quantum and classical Fisher information in such a setup, elucidating as well a straightforward measurement approach that nearly attains the maximum precision permissible under the Cramér-Rao bound. We showcase these results by focusing on the squeezing Hamiltonian, which characterizes the thermodynamic limit of Dicke and Lipkin-Meshkov-Glick Hamiltonians.
\end{abstract}
\date{\today}
\maketitle

%%%%%%%%%%%%%%%%%%%%%%%%%%%%%%%%%%%%%%%%%%%%%%%%%%%%%%%%%%%%%%%%%%%%%%%%%%%%
%%%%%%%%%%%%%%%%%%%%%%%%%%%%%%  INTRODUCTION  %%%%%%%%%%%%%%%%%%%%%%%%%%%%%%
%%%%%%%%%%%%%%%%%%%%%%%%%%%%%%%%%%%%%%%%%%%%%%%%%%%%%%%%%%%%%%%%%%%%%%%%%%%%

\emph{Introduction.}---Quantum metrology is a cornerstone of quantum technologies~\cite{QTECH2003milburn,QTECT2018roadmap} aimed at leveraging quantum phenomena for precision measurements, surpassing the limits imposed by classical physics~\cite{QM2006Llyod,QM2011Lloyd}. A key example involves exploiting quantum entanglement in systems composed of $N$ particles to overcome the statistical $\sqrt{N}$ enhancement (known as the standard quantum limit) and reaching the Heisenberg limit, where sensitivity scales linearly with the number of particles~\cite{pirandola2018rmp,QM2018rmp,QM2020Sciarrino}. In the simplest case, quantum correlations are generated within a system subjected to an unknown perturbation to be precisely determined. In the generic case of Ramsey interferometry~\cite{schmiedmayer2018ramsey,Schulte2020ramsey,kaubruegger2021ramsey,pedrozo2020entanglement,hotter2023ramsey,bohr2023collectively} the unknown perturbation (parameter) could be the magnetic field responsible for Zeeman splitting of the atomic levels (magnetometry) or the frequency detuning of a reference laser driving the atoms (atomic clocks). This external force introduces a change in the system's state (phase) depending on the value of the unknown parameter. Therefore, performing an appropriate measurement allows one to detect the change and decode the information about the unknown parameter concealed in the measurement outcomes. 

\begin{figure}[htb!]
    \centering
    \includegraphics[width=\columnwidth]{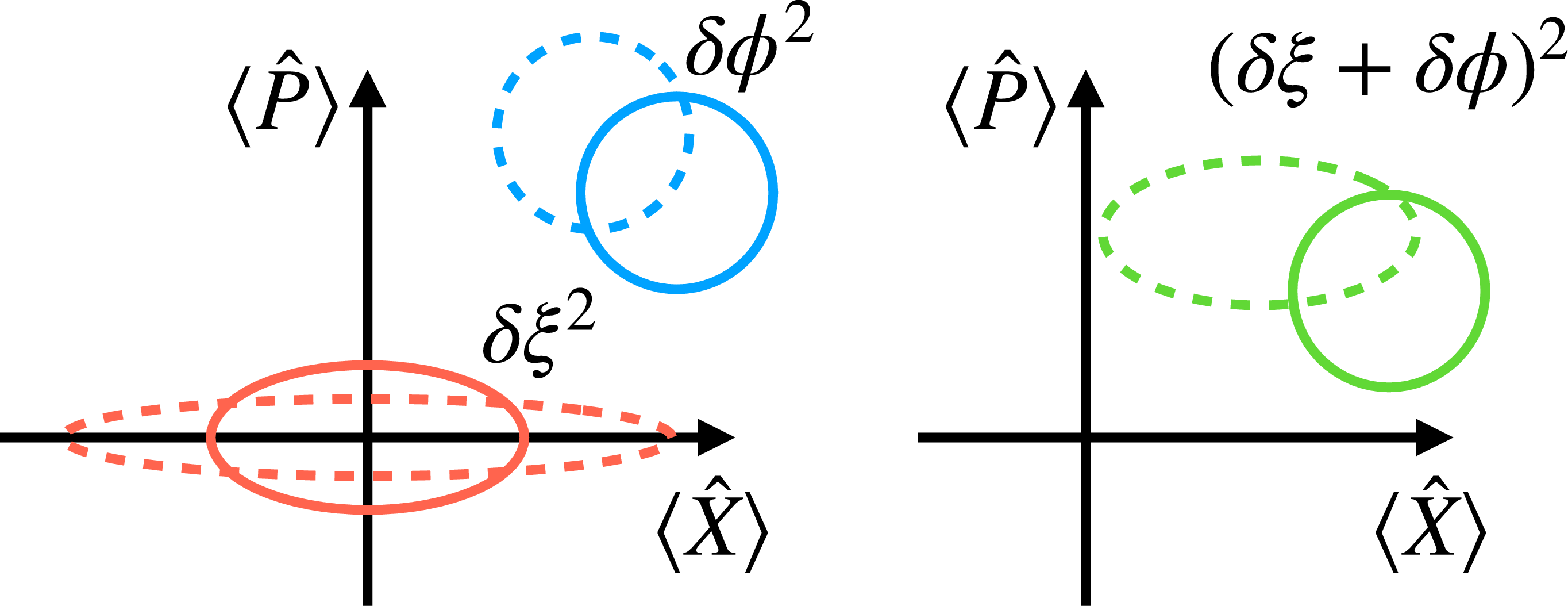}
    \caption{In conventional metrology (blue color), the information about the unknown parameter is stored in the accumulated phase $\delta \phi$. In critical metrology (red color), the information about the unknown parameter is stored in the squeezing parameter $\delta \xi$ (accumulated squeezing). By combining these two approaches (green color) it is possible to store the information in both the phase (position) and squeezing (shape) of the quantum state.}
    \label{fig:scheme}
\end{figure}

Unfortunately, creating and maintaining entanglement is technically very challenging owing to its inherent fragility~\cite{3H2009entanglement}. Furthermore, highly entangled states often require extensive detection schemes involving measuring complex many-body correlations~\cite{toth2009entdet} to harness their measurement power. For these reasons, quantum-enhanced measurements remain primarily in the domain of proof-of-principle experiments~\cite{demko2012elusive}. Nonetheless, significant theoretical and experimental efforts have emerged to overcome these challenges, leading to innovative methods and techniques.~\cite{ostermann2013protected,smith2016approachingHL,dur2016largeQFI,szigeti2017interactionbased, colombo2022satin, colombo2022entanglement, li2023improving}. A novel and promising approach here is critical metrology~\cite{zanardi2008quantumcriticliatresource,paris2014cqmLMG,garbe2020criticalmetrology,mihailescu2023multiparameter}, which relies on concurrently generating quantum correlations while exposing the system to the influence of a force of unknown strength~\cite{garbe2020criticalmetrology,Gietka2021adiabaticcritical,braak2022cqmnonlinearqr}. This differs from the conventional sequential approach of first creating a critical state and then subjecting it to an unknown perturbation~\cite{frerot2018criticalquantum}. In critical metrology, the system typically remains in its instantaneous ground state as the control parameter is adiabatically adjusted close to a critical value. Consequently, critical metrology protocols exhibit resilience to decoherence, albeit at the cost of the time required to complete the adiabatic ramp, a phenomenon known as critical slowing down. Recent findings challenge, however, the need for adiabatic ramps in critical metrology. Instead, they propose that protocols can achieve significantly improved performance by opting for a sudden quench rather than an adiabatic one. This insight has opened new avenues for advances in critical metrology methods~\cite{cai2021dynamicframeworkcqm,Gietka2021inverted,Gietka2022understanding,Garbe2022fromheisenbergtoKZ}.

The conventional practice of adhering to the instantaneous ground state via adiabatic dynamics confines the storage of information regarding an unknown parameter solely to the fast changing properties of the ground state near a critical point. However, when the initial state is not the ground state or another eigenstate of the system, the adiabatic dynamics also results in the encoding of some information about the unknown parameter in the accumulated phase differences between the contributing eigenstates (refer to Fig.~\ref{fig:scheme} for a schematic representation). 

In the following, we will present the generalized approach to combine these two metrological approaches and apply it to the example of a harmonic oscillator coupled to an ensemble of two-level systems (Dicke model). Additionally, we show how a position or momentum (homodyne type) measurement nearly saturates the combined Cram\'er-Rao bound for the corresponding measurement accuracy.

\emph{Combining critical and quantum metrology.}---The fusion of critical and conventional metrology becomes evident when examining the derivative of the wave function with respect to an unknown parameter, denoted as $\Omega$
\begin{align}\label{eq:derpsi}
     |\partial_\Omega \psi (\Omega) \rangle = & \partial_\Omega \left( \sum_{n=0}^{\infty}c_n(\Omega)|n(\Omega)  
    \rangle\right)=\\ = & \sum_{n=0}^{\infty}\left[\partial_\Omega c_n(\Omega)\right]|n(\Omega)\rangle + \sum_{n=0}^{\infty}c_n(\Omega)\left[\partial_\Omega|n(\Omega) \rangle\right],\nonumber
\end{align}
{where $\sum_{n=0}^{\infty}\left[\partial_\Omega c_n(\Omega)\right]|n(\Omega)\rangle$ represents the phase changes (conventional quantum metrology) and $\sum_{n=0}^{\infty}c_n(\Omega)\left[\partial_\Omega|n(\Omega) \rangle\right]$ represents the wavefunction's shape changes (critical metrology). This derivative can be used to calculate the quantum Fisher information $\mathcal{I}_\Omega = 4\left[\langle \partial_\Omega \psi|\partial_\Omega \psi\rangle -|\langle \psi|\partial_\Omega \psi\rangle |^2\right]$, which sets the %optimal uncertainty of estimating an unknown parameter 
minimal uncertainty to estimate an unknown parameter through the Cram\'er-Rao bound $\Delta\Omega \ge 1/\sqrt{\mathcal{I}_\Omega}$~\cite{caves1994QFI}. Calculating the corresponding expression of the quantum Fisher information}
yields (for %the sake of clarity 
readability we drop the dependence on $\Omega$)
\begin{align}
    \mathcal{I}_\Omega =& \, 4 \sum_n \left|\partial_\Omega c_n\right|^2 - 4\left|\sum_n c_n^*\partial_\Omega c_n \right| ^2 \\
    &+ 4 \sum_{n,m}  c_m c_n^*\langle \partial_\Omega n|\partial_\Omega m\rangle - \left|\sum_{n,m}c_m c_n^*\langle n| \partial_\Omega m\rangle\right| ^2 \\
    \begin{split}
    &+ 4 \sum_{n,m}c_n^* \partial_\Omega c_m\langle \partial_\Omega n | m \rangle + c.c. \\
    &- 4 \sum_{n,m,k} c_n^* c_m c_k^*\partial_\Omega c_k\langle \partial_\Omega n | m \rangle +c.c.
    \end{split}
\end{align}
where $c.c.$ stands for the complex conjugate. The first line in the above equation is related to conventional metrology ($\mathcal{I}^\phi$), the second line to critical metrology ($\mathcal{I}^\xi$), and the last two lines describe the interference between critical and quantum metrology ($\mathcal{I}^I$). {The total quantum Fisher information is the sum of the three parts $\mathcal{I}_\Omega = \mathcal{I}^\phi + \mathcal{I}^\xi + \mathcal{I}^I$}.

The quantum Fisher information accurately defines the fundamental limit of uncertainty, assuming an optimal measurement has been executed to extract information regarding the unknown parameter. However, in some instances, conducting such measurements can be exceedingly intricate and may surpass the capabilities of current experimental techniques. Consequently, we have to consider the classical Fisher information, which factors in a well-defined measurement strategy under practical constraints. The classical Fisher information can be defined using the conditional probability $p(\chi|\Omega)$ of observing an outcome labeled by $\chi$ given $\Omega$ in the following way
\begin{align}
    \mathcal{F}_\Omega = \sum_{\chi} \frac{1}{p(\chi|\Omega)}\left(\frac{\partial p(\chi|\Omega)}{\partial \Omega} \right)^2.
\end{align}

Numerous relevant physical systems and quantum technologies hinge on the interplay among Gaussian states, Gaussian operations, and Gaussian measurements~\cite{gaussian2021lloyd,brask2022gaussian}. This apparent limitation to the Gaussian realm, however, offers several advantages. It facilitates the utilization of straightforward analytical tools on the theoretical front and readily accessible components for implementing Gaussian processes in laboratory experiments. If the probability $p(\chi|\Omega)$ is Gaussian (characterized by the first and the second moment only), the classical Fisher information can be simplified to \cite{kay1993fundamentals}
\begin{align} \label{eq:error_prop}
    \mathcal{F}_\Omega = \frac{\left(\partial_\Omega \langle \hat O \rangle\right)^2}{\Delta^2 \hat O} + \frac{1}{2}\frac{\left(\partial_\Omega \Delta^2 \hat O\right)^2}{\left(\Delta^2\hat O\right)^2},
\end{align}
where $\langle \hat O \rangle = \sum_\chi \chi p(\chi|\Omega)$ is the mean of the distribution and $\Delta^2\hat O$ is its variance. 
{If $\hat{O}$ represents a quadrature or collective spin operator then the second term is non-zero only if the quantum state changes its shape. This means it only corresponds to the critical metrology contribution to the classical Fisher information. The first term, on the other hand, contains the conventional and interference term, which means it disappears for solely critical metrology protocols.} 

%%%%%%%%%%%%%%%%%%%%%%%%%%%%%%%%%%%%%%%%%%%%%%%%%%%%%%%%%%%%%%%%%%%%%%%%%%%%
%%%%%%%%%%%%%%%%%%%%%%%%%%%%%%  DICKE MODEL  %%%%%%%%%%%%%%%%%%%%%%%%%%%%%%%
%%%%%%%%%%%%%%%%%%%%%%%%%%%%%%%%%%%%%%%%%%%%%%%%%%%%%%%%%%%%%%%%%%%%%%%%%%%%

\emph{Dicke model.}---The Dicke model describes the interaction between a quantized harmonic oscillator and a collection of two-level systems~\cite{barry2011dickemodelrevisited}. Notably, the Dicke model can be simulated using various physical platforms~\cite{solano2014simulationDicke,engels2014DickeSOCBEC,dickesimulator2018amr,lamata2018dickeIONS}. Its Hamiltonian can be expressed as ($\hbar = 1$)
\begin{align}
    \hat H = \omega \hat a^\dagger \hat a + \frac{\Omega}{2}\sum_{i=1}^N \hat \sigma_z^i + \frac{g}{2 \sqrt{N}}\left(\hat a^\dagger + \hat a\right)\sum_{i=1}^N \hat \sigma_x^i,
\end{align}
where $\omega$ is the frequency of the harmonic mode created by $\hat a^\dagger$ and annihilated by $\hat a$, and $\Omega$ is the frequency of a single two-level system described by the set of Pauli matrices $\hat \sigma_i$ with $i=x,y,z$. The two sub-systems are coupled with strength $g$. For $\Omega \gg \omega$, the dynamics of the spins can be eliminated \cite{Helmut2013rmpcavity}, and provided the coupling is not greater than the critical coupling $g_c\equiv \sqrt{\Omega \omega}$, the effective Hamiltonian becomes~\cite{feshke2012dickemodelQPT,oriol2022unstable}
\begin{align}\label{eq:Heff}
       \hat H \approx \omega \hat a^\dagger \hat a - \frac{g^2}{4\Omega}\left(\hat a^\dagger + \hat a\right)^2,
\end{align}
which is a squeezing Hamiltonian typically studied in the context of critical metrology~\cite{paris2014cqmLMG,garbe2020criticalmetrology,e2021teklu,cai2021dynamicframeworkcqm,Gietka2021adiabaticcritical,Gietka2022understanding,plenio2022PRX,gietka2022criticalspeedup,gietka2022commetrology}. In the following, we assume that we always stay in regimes where the effective Hamiltonian~\eqref{eq:Heff} is valid. Note that the above Hamiltonian also describes the Lipkin-Meshkov-Glick model in the thermodynamic limit.
% In order to create an intuitive picture, we introduce abstract position and momentum operators and rewrite the Hamiltonian in the form of a harmonic oscillator with a unit mass
% %
% \begin{align}\label{eq:abstract}
%     \hat H = \frac{\hat p^2}{2} + \frac{\omega^2}{2}\left(1-\frac{g^2}{g_c^2}\right)\hat x^2,
% \end{align}
% %
% with a frequency $\omega\sqrt{1-g^2/g_c^2}$, where $g_c \equiv \sqrt{\omega \Omega}$. In this description it becomes clear %It becomes clear now 
% that by increasing the coupling parameter $g$ the abstract harmonic oscillator \emph{opens} which creates squeezing with respect to the unperturbed harmonic oscillator $\omega \hat a^\dagger \hat a \equiv \frac{\hat p^2}{2} + \frac{\omega^2}{2}\hat x^2$. 
The eigenspectrum of the Hamiltonian~\eqref{eq:Heff} is %can be easily found
\begin{align}
    |\psi_n\rangle = \exp\bigg\{\frac{1}{2}\left(\xi^*\hat a^2-\xi\hat a^{\dagger2}\right)\bigg\}|n\rangle,
\end{align} 
where $\xi = \frac{1}{4} \ln\{1-g^2/g_c^2\}$ is the squeezing parameter and $|n\rangle$ are Fock states generated by $\hat a^\dagger$ acting on the vacuum $|0\rangle$. These states can be used to construct an arbitrary wave function and calculate the Fisher information. % which is the key figure in quantum metrology as it is related to the uncertainty of estimating an unknown parameter through the Cram\'er-Rao bound $\Delta\Omega \ge 1/\sqrt{\mathcal{I}_\Omega}$, where $\mathcal{I}_\Omega$ is the quantum Fisher information for the unknown parameter $\Omega$.

%%%%%%%%%%%%%%%%%%%%%%%%%%%%%%%%%%%%%%%%%%%%%%%%%%%%%%%%%%%%%%%%%%%%%%%%%%%%
%%%%%%%%%%%%%%%%%%%%%%%%%%%%%  QUANTUM FISHER  %%%%%%%%%%%%%%%%%%%%%%%%%%%%%
%%%%%%%%%%%%%%%%%%%%%%%%%%%%%%%%%%%%%%%%%%%%%%%%%%%%%%%%%%%%%%%%%%%%%%%%%%%%

In order to increase the dependence of the wave function on the unknown parameter, instead of using the ground state (critical metrology), we first displace the initial state creating a coherent state characterized by $\alpha \equiv |\alpha|e^{i \arg \alpha}$. Subsequently, we adiabatically increase the coupling (control) parameter from 0 towards the critical point $g_c$. According to the adiabatic theorem the final state is given by~\cite{gietka2023unique}
\begin{align}
    |\psi(t)\rangle = e^{-\frac{|\alpha|^2}2} \sum_{n=0}^\infty e^{-i\int_{0}^{t}E_n(t^\prime)\mathrm{d}t^\prime} e^{i \gamma_n} \frac{\alpha^n}{\sqrt{n!}}|n(t)\rangle,
\end{align}
where $E_n(t) = n\omega\sqrt{1-g(t)^2/g_c^2}$ is the $n$-th instantaneous eigenenergy of Hamiltonian~\eqref{eq:Heff} and $\gamma_n = \int_{0}^g \langle n(g^\prime)| \partial_{g^\prime}|n(g^\prime)\rangle \mathrm{d}g^\prime $ is the geometric (Berry) phase~\cite{berry1984adiabaticphase}. It can be shown that a ramp~\cite{garbe2020criticalmetrology}
\begin{align}
    g(t) = 2 g_c \frac{\sqrt{\gamma \omega t(\gamma \omega t +1)}}{2\gamma \omega t +1},
\end{align}
with $\gamma \ll 1$ satisfies the adiabatic condition. Assuming $g(T)\equiv g_f\sim g_c$, the total time of the evolution becomes
\begin{align}
    T = \frac{1}{2\gamma \omega\sqrt{1-g_f^2/g_c^2}},
\end{align}
where $g_f$ is the final value of the coupling. 

\begin{figure}[htb!]
    \centering
    \includegraphics[width=\columnwidth]{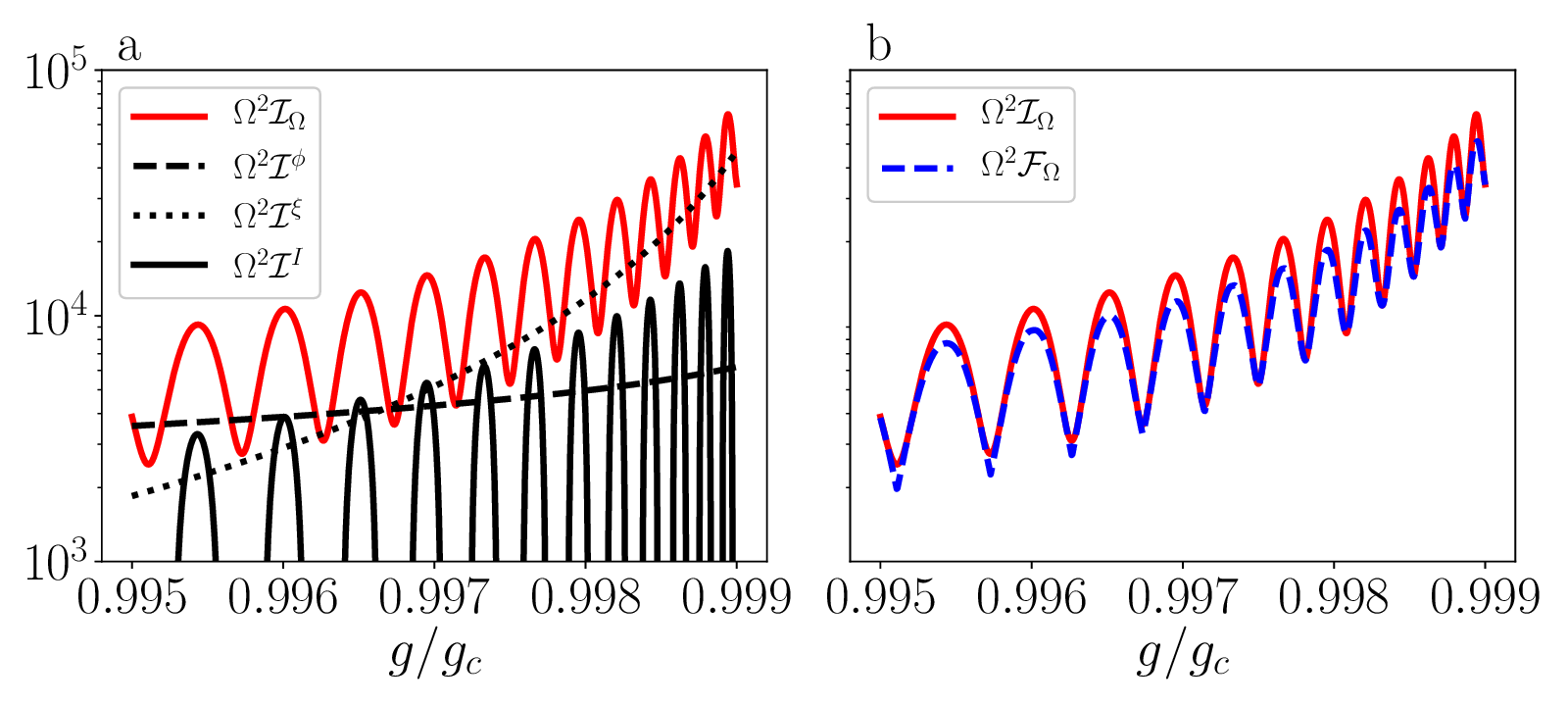}
    \caption{Quantum and classical Fisher information. (a) depicts a comparison among the contributions to the quantum Fisher information for an initial state with $\alpha = 0.5e^{-0.3i}$ as a function of $g/g_c$. (b) depicts quantum (solid red line) and quadrature-optimized classical Fisher information (dashed blue line). For these simulations, we set $\Omega/\omega = 200$.}
    \label{fig:fig1}
\end{figure}

We can now proceed to calculate the quantum Fisher information. This can be done by using a derivative of the instantaneous wave function with respect to the unknown parameter
\begin{align}
\begin{split}
 |\partial_\Omega \psi \rangle =& e^{-\frac{|\alpha|^2}2} \sum_{n=0}^\infty \left[\partial_\Omega e^{-i\int_{0}^{T}E_n(t)\mathrm{d}t}\right] \frac{\alpha^n}{\sqrt{n!}}|n(T)\rangle \\
 &+ e^{-\frac{|\alpha|^2}2} \sum_{n=0}^\infty e^{-i\int_{0}^{T}E_n(t)\mathrm{d}t} \frac{\alpha^n}{\sqrt{n!}}\partial_\Omega|n(T)\rangle.
\end{split}
\end{align}
Note that the Berry phase is zero because the eigenstates $|n(T)\rangle$ are real~\cite{zanardi2007geometrictensors}. In order to calculate the quantum Fisher information, we simplify the above expression. Let us have a look at the first term containing
\begin{align}
    \partial_\Omega e^{-i\int_{0}^{T}E_n(t)\mathrm{d}t} = \partial_\Omega e^{-i n \int_{0}^{T}E_0(t)\mathrm{d}t},
\end{align}
where $E_0(t)$ is the instantaneous energy gap between two neighboring energy levels (this is only true for a harmonic oscillator). By plugging the adiabatic ramp and the expression for the final time, the accumulated phase becomes (assuming $g_f\sim g_c$)
\begin{align}
 \phi = {\int_{0}^{T}E_0(t)\mathrm{d}t} = \frac{ g_f \log \left[\frac{1}{\sqrt{1-{g_f^2}/{{g_c^2}}}}+1\right]}{2 \gamma  \sqrt{\Omega \omega}}.
\end{align}
This allows us to conveniently rewrite the final state as
\begin{align}\label{eq:stateT}
    |\psi(T)\rangle = \hat S(\xi)\hat D(\alpha e^{-i \phi})|0\rangle.
\end{align}
By acting with the derivative operator on the above wave function, we get
\begin{align}
    |\partial_\Omega \psi(T)\rangle  = & \frac{\left( \hat a^2 - \hat a^{\dagger2}  \right)\hat S(\xi)\hat D(\alpha e^{-i \phi})|0\rangle}{8\Omega({g_c^2}/{g^2}-1)} \\ &+
    \frac{i \phi \alpha e^{-i \phi} }{2\Omega} \hat S(\xi)\hat a^\dagger \hat D(\alpha e^{-i \phi})|0\rangle,
\end{align}
which we use to calculate the Fisher information. After some straightforward algebra, the contribution from the dynamical phase (conventional metrology) becomes 
\begin{align}
    \mathcal{I}^\phi = \frac{ \phi^2}{\Omega^2}|\alpha|^2,
\end{align}
and the contribution from the adiabatic change of the eigenstates (critical metrology) becomes 
\begin{align}\label{eq:qfiAclosed}
    \mathcal{I}^\xi = \frac{1+2|\alpha|^2}{8 \Omega^2\left(1-{g_c^2}/{g^2}\right)^2} =  \frac{g^4}{g_c^4}\frac{1+2|\alpha|^2}{8 \Omega^2e^{8\xi}}.
\end{align}
Interestingly, the contribution from the critical metrology is amplified by the use of excitations represented by $|\alpha|^2$. For the term originating from the interference between conventional and critical metrology, we obtain 
\begin{align}\label{eq:qfiint1}
   \mathcal{I}^I =  \frac{|\alpha|^2  \phi  \sin [2(\arg \alpha-\phi) ]}{ \Omega ^2 \left(1-{g_c^2}/{g^2}\right)},
\end{align}
which can be negative (destructive) or positive (constructive) depending on the final coupling $g_f$. The three components of the quantum Fisher information are shown in Fig.~\ref{fig:fig1}. Specifically, we choose the simulation parameters to highlight the moment when the contribution from critical metrology overcomes the contribution from conventional metrology. 

It is worth noting that similar calculations can be performed treating $\omega$ or $\Omega$ as a control parameter and defining $\omega_c$ or $\Omega_c$ as critical frequencies. For instance, in spin-orbit coupled quantum gases, $\Omega$ is typically the control parameter~\cite{Busch2016SOCBEC}. Then, treating $\omega$ as unknown and performing analogous calculations, it is straightforward to show the following relation
\begin{align}
    \frac{\mathcal{I}_\omega}{\mathcal{I}_\Omega} = \frac{\Omega^2}{\omega^2}.
\end{align}

%%%%%%%%%%%%%%%%%%%%%%%%%%%%%%%%%%%%%%%%%%%%%%%%%%%%%%%%%%%%%%%%%%%%%%%%%%%%
%%%%%%%%%%%%%%%%%%%%%%%%%%%  CLASSICAL FISHER  %%%%%%%%%%%%%%%%%%%%%%%%%%%%%
%%%%%%%%%%%%%%%%%%%%%%%%%%%%%%%%%%%%%%%%%%%%%%%%%%%%%%%%%%%%%%%%%%%%%%%%%%%%

\emph{Optimal measurements.}---Since the state from Eq.~\eqref{eq:stateT} is Gaussian, we can expect that measuring an appropriate quadrature will yield a lot of information about the unknown parameter and nearly saturate the Cram\'er-Rao bound. The generalized quadrature is defined as $\hat Q = \hat X \cos \theta + \hat P \sin \theta$, where $\hat X = (\hat a + \hat a^\dagger)/2$ and $\hat P = (\hat a - \hat a^\dagger)/2i$. For the final state, the $\hat X$ and $\hat P$ quadrature average values are
\begin{align}\label{eq:mftraj}
\begin{split}
\langle \hat X \rangle &= |\alpha|\cos(\phi-\arg \alpha) \exp(-\xi) ,\\
\langle \hat P \rangle &=-|\alpha|\sin(\phi-\arg \alpha) \exp(\xi).
\end{split}
\end{align}
The generalized quadrature variance is
\begin{align}
    \Delta^2 \hat Q =  \frac{\exp(-2\xi)}{4} \cos^2 \theta +\frac{\exp(2\xi)}{4}  \sin^2 \theta.
\end{align}
Using the Fisher information formula \eqref{eq:error_prop}, %it is straightforward to 
we can calculate the classical Fisher information for arbitrary parameters. For the sake of brevity, however, we only provide the expressions for the $\hat X$ and $\hat P$ quadratures. Assuming $\arg \alpha = 0$, we get
\begin{align}
\begin{split}
    \mathcal{F}_\Omega^{\hat X} =& |\alpha|^2\frac{\left(\frac{g^2}{g_c^2}\cos(\phi)e^{-4\xi} -2\phi \sin(\phi) \right)^2}{4\Omega^2} +  \frac{g^4}{g_c^4}\frac{1}{8 \Omega^2e^{8\xi}},\\
   \mathcal{F}_\Omega^{\hat P} =& |\alpha|^2\frac{\left(\frac{g^2}{g_c^2}\sin(\phi)e^{-4\xi} -2\phi \cos(\phi) \right)^2}{4\Omega^2} +  \frac{g^4}{g_c^4}\frac{1}{8 \Omega^2e^{8\xi}}.
   \end{split}
\end{align}
The above classical Fisher information saturates the Cram\'er-Rao bound whenever the state is aligned with the $ \hat X$ or $\hat P$ quadrature, respectively. Although these two measurements saturate the Cram\'er-Rao bound, they saturate it only if the interference term in the quantum Fisher information is equal to 0. Unfortunately, numerical calculations indicate that once the interference term is maximal, the Cram\'er-Rao bound cannot be saturated by quadrature measurements as illustrated in Fig.~\ref{fig:fig1}b, where we compare the quantum Fisher information and the quadrature-optimized classical Fisher information. This means that different (non-commuting) quadrature measurements are required to extract the maximum amount of information about the unknown parameter from the average and the variance [see Eq.~\eqref{eq:error_prop}] once the quantum state is not aligned with the $\hat X$ or $\hat P$ quadratures. Nevertheless, even when the interference term is maximal, a substantial amount of information can be still extracted from appropriate quadrature measurement nearly saturating the Carm\'er-Rao bound (see Fig.~\ref{fig:fig1}b).

%%%%%%%%%%%%%%%%%%%%%%%%%%%%%%%%%%%%%%%%%%%%%%%%%%%%%%%%%%%%%%%%%%%%%%%%%%%%
%%%%%%%%%%%%%%%%%%%%%%%%%%  DRIVEN-DISSIPATIVE  %%%%%%%%%%%%%%%%%%%%%%%%%%%%
%%%%%%%%%%%%%%%%%%%%%%%%%%%%%%%%%%%%%%%%%%%%%%%%%%%%%%%%%%%%%%%%%%%%%%%%%%%%

\emph{Driven-dissipative case.}---A driven-dissipative system is a type of physical system that is subject to both external driving forces and dissipative processes. A primary example is a laser-pumped (driven) optical cavity that loses photons (dissipation) through imperfect mirrors. These systems are of particular interest because they can exhibit complex and fascinating behavior, including the emergence of non-equilibrium phenomena and non-equilibrium phase transitions~\cite{sieberer2013drivendissiaptve,esslinger2013drivedisispaiveDICKE} which can also be harnessed in quantum metrology~\cite{Pavlov_2023_drivendisiipative}, {in particular, for joint estimation of loss and nonlinearity in Kerr resonators~\cite{teklu2023DDKR}.}

Although the dissipation is typically considered a metrological disadvantage, in the driven-dissipative case, we can expect that the ramp can be much faster as the dissipation will eventually %\emph{kill} 
destroy non-adiabatic excitations. In order to account for the dissipation, we use the master equation in the Lindblad form
\begin{align}
\begin{split}
\mathcal{L}[\hat \rho] = &  -i\left[\hat H + \eta (\hat c e^{i \omega_d t} + \hat c^\dagger e^{-i \omega_d t}), \hat \rho \right]\\&   +\kappa\left(\hat c \hat \rho \hat c^\dagger -\frac{1}{2}\big\{\hat c^\dagger\hat c, \hat \rho \big\}\right), 
\end{split}
\end{align}
{where $\hat H$ is the Hamiltonian from Eq.~\eqref{eq:Heff}, $\kappa$ denotes excitation losses (dissipation), and $\eta$ is the strength of the drive with frequency $\omega_d$}. Note that in the strong coupling regime close to the critical point $\hat a$ is no longer the correct jump or drive operator, and we have to use modified jump operators $\hat c$~\cite{nori2020ultrastrong,dissipation2011blais,rabl2018dissipativeQRM,Cattaneo_2019,ultrastrong2019solano,gietka2023unique}. After reaching the steady state solution (equilibrium state), the system is characterized by 
\begin{align}\label{eq:ddstate}
    |\psi (t) \rangle =\hat S(\xi)\hat D(|\tilde \alpha|\exp[-i(\omega_d t +\varphi)])|0\rangle,
    %e^{-\frac{|\tilde \alpha|^2}{2}} \sum_{n=0}^\infty e^{-i n\left( \omega_d t + \phi \right)}\frac{|\tilde\alpha|^n}{\sqrt{n!}}e^{\frac{1}{2}\left(\xi^* \hat a^2-\xi \hat a^{\dagger2} \right)}|n \rangle \\
\end{align}
where $\tilde \omega = \omega \sqrt{1-g^2/g_c^2} = \omega e^{2\xi}$ is the resonance frequency, $|\tilde \alpha| = 2\sqrt{\eta^2/\left(\kappa^2 +4(\tilde \omega - \omega_d)^2\right)} $ is a parameter dependent amplitude, and $\varphi = \arctan [\frac{\kappa}{2(\tilde \omega -\omega_d)} ]$ is the phase difference between the drive and the response of the system. Note that the state from Eq.~\eqref{eq:ddstate} is pure which is a consequence of eliminating the spin degree of freedom from the equations. In a more general case when the spin degree of freedom cannot be eliminated from the description, the excitations of the spin will be entangled with the excitations of the harmonic oscillator. Consequently the driven-dissipative steady-state will be described by a density matrix for which an analytic description of different Fisher information contributions is usually nontrivial. Furthermore, for a driven-dissipative harmonic oscillator, the system will oscillate with the frequency of the driving force; therefore, the phase and thus the quantum Fisher information will not grow with time. We interpret this as a trade-off between reaching quickly the critical steady state and increasing the Fisher information. 

\begin{figure}[htb!]
    \centering
    \includegraphics[width=\columnwidth]{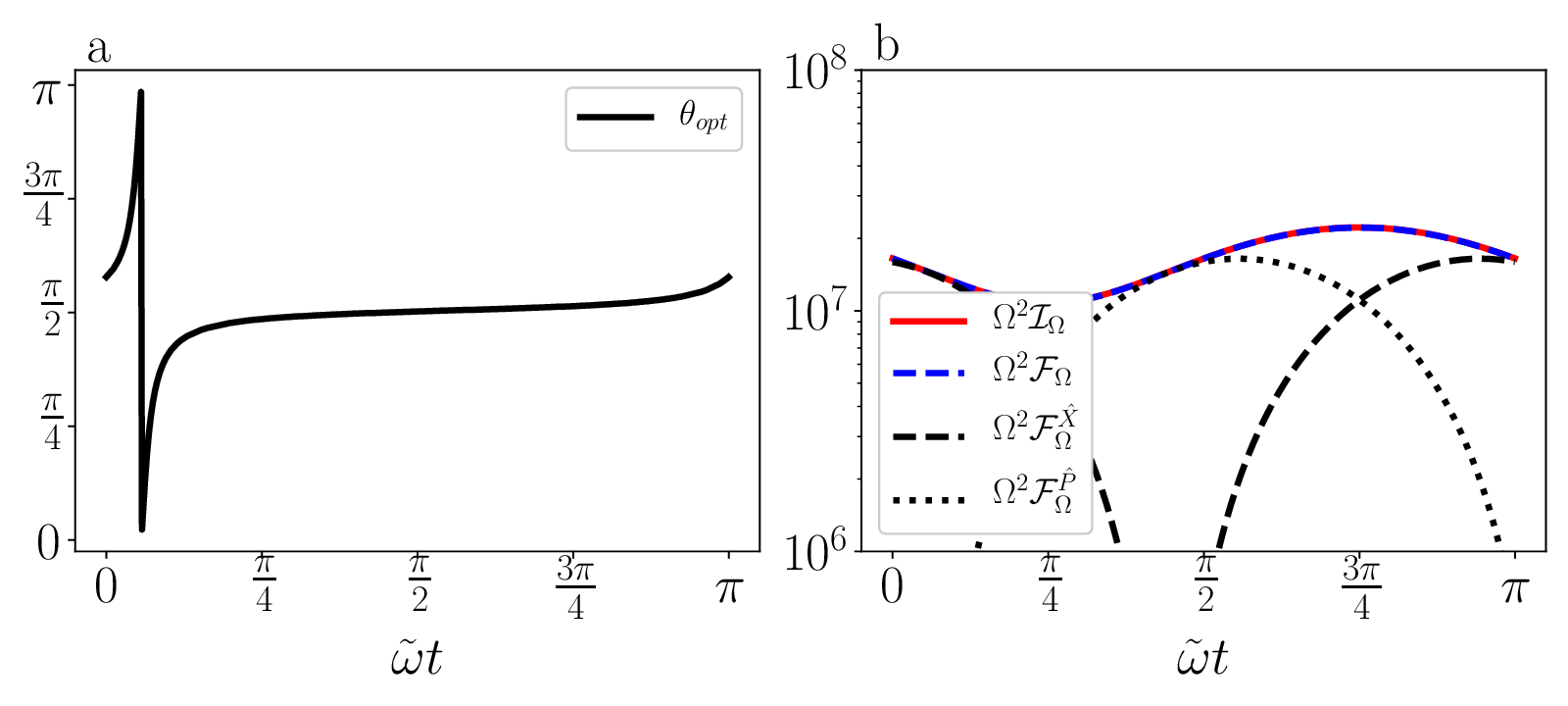}
    \caption{Quantum and classical Fisher information in the driven-dissipative case. (a) depicts the optimal quadrature angle as a function of time. (b) depicts quantum (solid red line) and classical Fisher information calculated using measurement of $\hat X$ (dashed line), $\hat P$ (dotted line), and optimal quadrature (dashed blue line). The parameters for these simulations are set to $\Omega/\omega =100.5$, $\eta/\omega =8$, $\kappa/\omega =1$, $g/g_c = 0.999$.}
    \label{fig:fig2}
\end{figure}

Following a similar derivation to that one in the previous section, the quantum Fisher information components for the resonant driving $\omega_d \approx \tilde \omega$ become
\begin{align}
    \mathcal{I}^\phi & = \frac{16 g^4 \eta^2}{(1-g^2/g_c^2)\Omega^4 \kappa^4},\\ \label{eq:IAdd}
    \mathcal{I}^\xi & = \frac{1+8\frac{\eta^2}{\kappa^2}}{8 \Omega^2\left(1-{g_c^2}/{g^2}\right)^2} ,\\
    \mathcal{I}^I & =  \frac{8 g^4 \eta^2 \sin\left[2(\omega_d t + \varphi) \right]}{g_c^2 \kappa^3 (1-g^2/g_c^2)^{3/2} \Omega^3},
\end{align}
and are analogous to the quantum Fisher information components in the isolated system dynamics. The quantum Fisher information in the driven-dissipative case is shown in Fig.~\ref{fig:fig2}b, where we compare it with the classical Fisher information calculated for various measurements.

%%%%%%%%%%%%%%%%%%%%%%%%%%%%%%%%%%%%%%%%%%%%%%%%%%%%%%%%%%%%%%%%%%%%%%%%%%%%
%%%%%%%%%%%%%%%%%%%%%%%%%%%% CLASSICAL FISHER  %%%%%%%%%%%%%%%%%%%%%%%%%%%%%
%%%%%%%%%%%%%%%%%%%%%%%%%%%%%%%%%%%%%%%%%%%%%%%%%%%%%%%%%%%%%%%%%%%%%%%%%%%%

In analogy to closed system dynamics, we also consider measurement of quadratures for quantum estimation. For the steady state, it is straightforward to calculate the means
\begin{align}
    \langle \hat X \rangle  &=  |\tilde \alpha| \cos\left( \omega_d t + \varphi\right) \exp(-\xi), \\
    %\Delta^2 \hat X &= \frac{\exp(-2\xi)}{4}, \\
    \langle \hat P \rangle &= - |\tilde \alpha| \sin\left( \omega_d t + \varphi\right) \exp(\xi) ,
    %\Delta^2 \hat P &=  \frac{\exp(2\xi)}{4}.
\end{align}
and the variance
\begin{align}
    \Delta^2\hat Q =  \frac{\exp(-2\xi)}{4} \cos^2 \theta +\frac{\exp(2\xi)}{4}  \sin^2 \theta.
\end{align}
Using the classical Fisher information formula and assuming resonant condition $\tilde \omega = \omega_d$, one gets
%\begin{widetext}
\begin{align}
\begin{split}
    \mathcal{F}_\Omega^{\hat X} = \frac{\left[\eta \kappa \cos (\omega_d t) - 4 \eta \tilde \omega \sin (\omega_d t)\right]^2}{\kappa^4 \Omega^2(1-g_c^2/g^2)^2}  +  \frac{g^4}{g_c^4}\frac{1}{8 \Omega^2e^{8\xi}},\\ 
   \mathcal{F}_\Omega^{\hat P} = \frac{\left[\eta \kappa \sin (\omega_d t) - 4 \eta \tilde \omega \cos (\omega_d t)\right]^2}{\kappa^4 \Omega^2(1-g_c^2/g^2)^2} +  \frac{g^4}{g_c^4}\frac{1}{8 \Omega^2e^{8\xi}},\\,
\end{split}
\end{align}
which saturates the Cram\'er-Rao bound whenever the state is aligned with $\hat X$ and $\hat P$ quadrature, respectively. Similarly as in the close system case, the maximum of the quantum Fisher information occurs once the interference term is maximal which happens whenever $\omega_d t = {\pi}/{4} + n\pi$ (assuming resonance condition $\varphi = 0$), and the appropriate quadrature measurement nearly saturates the Cram\'er-Rao bound (see Fig.~\ref{fig:fig2}b). Additionally, in Fig.~\ref{fig:fig2}a, we show the angle of the optimal quadrature measurement. Note that in Fig.~\ref{fig:fig2}b, it seems as if the Cram\'er-Rao bound can be saturated for an arbitrary time. In fact, for the simulation parameters we nearly saturate the Cram\'er-Rao bound and the difference between classical and quantum Fisher information is negligible.

%%%%%%%%%%%%%%%%%%%%%%%%%%%%%%%%%%%%%%%%%%%%%%%%%%%%%%%%%%%%%%%%%%%%%%%%%%%%
%%%%%%%%%%%%%%%%%%%%%%%%%%%%%%  CONCLUSIONS  %%%%%%%%%%%%%%%%%%%%%%%%%%%%%%%
%%%%%%%%%%%%%%%%%%%%%%%%%%%%%%%%%%%%%%%%%%%%%%%%%%%%%%%%%%%%%%%%%%%%%%%%%%%%

\emph{Conclusions.}---In this manuscript, we proposed to combine critical and quantum metrology into one effective protocol. This can be intuitively understood as storing the information about an unknown parameter both in the position of the wave function in the phase space and its shape. As a consequence, the quantum Fisher information attains an extra term related to the interference between critical and quantum metrology. We considered the case of closed system dynamics where the quantum Fisher information grows in time at the expense of adiabatic time evolution, and the case of driven-dissipative (open) dynamics where the quantum Fisher information does not grow in time but the adiabatic time evolution is not required. We also found a relatively simple measurement scheme that relies on measuring average values of quadratures and nearly saturates the Cram\'er-Rao bound.

Ideal candidates for experimental realization of combining critical and conventional metrology are systems where at least one (control) parameter can be tuned over a wide range of values. Examples include non-interacting spin-orbit coupled Bose-Einstein condensates~\cite{Busch2016SOCBEC,gietka2022commetrology} and spin-orbit coupled Fermi gases~\cite{gietka2023HSsoc} where the motional degree of freedom represents a harmonic oscillator and the spin degree of freedom represent a two-level system; quantum simulators realizing quantum Rabi model where the phonon mode represents a harmonic oscillator~\cite{duan2021qrbphonon}; and purely spin systems that can be mapped to the Lipkin-Meshkov-Glick Hamiltonian~\cite{Chen:09LMG,pengbo2017LMGSim,li2023improving}. In principle, light-matter systems~\cite{ultrastrong2019solano} could also be tested for combining critical and conventional metrology protocols. This would require, however, an extra step of turning off the light-matter interactions converting thus virtual excitations into real ones~\cite{giannelli2023detecting,unvelingvirtual2023nori}, which might constitute a serious obstruction.

Future plans include additional exploitation of the Berry Phase to store the information about the unknown parameter. In such a case, the quantum Fisher information could contain six terms. One related to conventional metrology, one related to critical metrology, one related to geometric metrology (Berry phase), and three interference terms. A promising candidate is an extension to the Dicke model which includes the term $\hat \sigma_y(\hat a^\dagger -\hat a)$ realizing squeezing in the orthogonal direction to $\hat \sigma_x(\hat a^\dagger + \hat a)$. Also using squeezed initial states might be an interesting direction for a future investigation.
%\kgc{drawbacks and comparisons...}

\begin{acknowledgments}
K.G. is pleased to acknowledge Hai-Long Shi, Lewis Ruks, Rafał Demkowicz-Dobrza\'nski, and Jan Kołody\'nski for fruitful discussions. Simulations were performed using the open-source \textsc{QuantumOptics.jl}~\cite{kramer2018quantumoptics} framework in \textsc{Julia}. This work was supported by the Lise-Meitner Fellowship M3304-N of the Austrian Science Fund (FWF).
\end{acknowledgments}

% \ch{General comments: Vladan scrambling paper (LMG experimental); sudden quench possible? - other paper; storyline - reorder sections; delete x and p part (model); }

%%%%%%%%%%%%%%%%%%%%%%%%%%%%%%%%%%%%%%%%%%%%%%%%%%%%%%%%%%%%%%%%%%%%%%%%%%%%
%%%%%%%%%%%%%%%%%%%%%%%%%%%%%%  BIBLIOGRAPHY %%%%%%%%%%%%%%%%%%%%%%%%%%%%%%%
%%%%%%%%%%%%%%%%%%%%%%%%%%%%%%%%%%%%%%%%%%%%%%%%%%%%%%%%%%%%%%%%%%%%%%%%%%%%

%\bibliography{bibliography.bib}


\end{document}